\documentclass[twocolumn,10pt]{article}

\usepackage{amsmath,amssymb,epsfig,cite,authblk,bbm}
\usepackage{color,hyperref}
\hypersetup{
    colorlinks,%
    citecolor=blue,%
    filecolor=blue,%
    linkcolor=black,%
    urlcolor=blue
}
\definecolor{bluecolor}{rgb}{0,0.,1.}

\definecolor{redcolor}{rgb}{.7,0.,0.}

\newcommand{\pr}[1]{\left( #1\right)}
\newcommand{\prr}[1]{\left[ #1 \right]}
\newcommand{\es}[1]{\begin{equation}\begin{split}#1\end{split}\end{equation}}

\newcommand{\R}{\mathbb{R}}

\newcommand{\br}{\boldsymbol{r}}

\newcommand{\bx}{\boldsymbol{x}}

\newcommand{\dd}{\textrm{d}}

\newcommand{\Ck}{\color{black}}

\newcommand{\bv}{\boldsymbol{v}}

\begin{document}
\title{Spatial networks with wireless applications}
\author[1]{Carl P. Dettmann} 
\author[1,2]{Orestis Georgiou}
\author[1]{Pete Pratt}
{\small
\affil[1]{School of Mathematics, University of Bristol, University Walk, Bristol, BS8 1TW, UK}
\affil[2]{Ultrahaptics, The West Wing, Glass Wharf, Bristol, BS2 0EL, UK}
}

\twocolumn[
  \begin{@twocolumnfalse}
    \maketitle
Many networks have nodes located in physical space, with links more common between
closely spaced pairs of nodes.  For example, the nodes could be wireless devices and
links communication channels in a wireless mesh network.  We describe recent work
involving such networks, considering effects due
to the geometry (convex, non-convex, and fractal), node distribution, distance-dependent 
link probability, mobility, directivity and interference.
\vspace*{20pt}
  \end{@twocolumnfalse}
]
\setcounter{tocdepth}{2}
\tableofcontents

\section{Introduction}
Many real networks have a spatial structure in that the nodes have locations,
and pairs of close nodes are more likely to be linked.  
This work will largely be concerned with applications to wireless communications, however much of the analysis is of far more general relevance, as well as being of theoretical interest. The wireless applications include mesh networks where information is relayed in a multihop fashion from node to node rather than directly to a central router or base station. Generally speaking, a mesh network models a collection of low power nodes (where long range links are unlikely) communicating to one another, for example the nodes can represent smart devices such as phones and laptops, and links represent a wireless channel. Many of the results assume that a node can receive and transmit data simultaneously (``full duplex").  Alternatively, it should be specified which nodes are acting as transmitters and receivers, leading to directed graphs. Interference, the effects due to unwanted signals from other transmitters, may need to be taken into account, and can be mathematically involved since the existence of a working link depends not only on the locations of the two nodes, but of all other transmitters. In Sec\ref{s:pp} we also discuss a range of different point processes which are used to model the fixed network architecture where the nodes model the location of base stations.

Even in the first paper on spatial networks~\cite{Gilbert61}, communication networks were the stated motivation.
The model considered
there, the random geometric graph, was first given this designation in Ref.~\cite{JAMS89}. In the meantime, there has been a substantial body of work in
probability~\cite{Penrose03} and communications~\cite{Haenggi12} with applications in many other fields~\cite{Bart11}.
 
A random geometric graph (RGG) comprises randomly located nodes, with links formed between mutually close pairs.  Mathematically, the locations of nodes are described by a point process $\Phi$, that is, a random set of points in a space $X$, and the concept of closeness by a distance function ${\cal D}:X\times X\to\mathbb{R}$. Most work on RGG use as the space $X$ either Euclidean space $\mathbb{R}^d$, the unit cube $[0,1]^d$, or the flat torus obtained by identifying opposite faces of the cube.  The latter is finite and homogeneous, that is, all points are equivalent.  We consider each of these mathematical ingredients in turn in the subsequent sections.

In the original Gilbert RGG model~\cite{Gilbert61}, nodes were distributed according to a uniform Poisson Point Process (see Sec.~\ref{s:ppp} below)
on $\mathbb{R}^2$ with links made between pairs of nodes within a fixed distance $r_0$. In Ref\cite{Gilbert61} the aim was to address percolation (see Sec.~\ref{s:perc})
with a random spatial structure and spatially dependent links; in essence this work gave  rise to the field of continuum percolation where there is no fixed underlying lattice structure. 
Many results for the RGG in $d\geq 2$ are reviewed in Walters~\cite{Walters11},
including on maximum and minimum degrees, cliques, percolation, (k)-connectivity, Hamiltonicity, chromatic number and coverage. 

For the application to wireless networks it largely remains  a balancing act between mathematical tractability and accuracy of the model.
It therefore makes sense to focus on particular network characteristics such as regions with high densities (a city shopping centre on a weekend), the fractal distribution of waypoints \cite{chen2018capacity, D17} or the bottle necks to connection probability which allow for tractable analysis that can capture the essence of the problem.

\section{Point processes}\label{s:pp}
\subsection{Stochastic Geometry}

Stochastic geometry is the study of random sets in space, most notably point processes, that is, random sets of individual points.
Initially, stochastic geometry was first used to further understanding in fields such as material science, astronomy and biology \cite{soshnikov2000determinantal,torquato2008point, Baccelli2010}.
Generally speaking, in stochastic geometry, point processes need not model just collections of points in some space, they can be used for more general
sets such as balls, lines, planes and fibres which are then mapped back into point processes using a suitable representation \cite{Kendall2010, Schmidt2014}.
Stochastic geometry is certainly not limited to the study of wireless networks; examples of other applications include material science (modelling of fibres),
astronomy (of which Olbers' paradox is a nice example), biology and ecology to model say forestry distributions and more recently in machine learning \cite{Kulesza2012}.

With this in mind, many authors have leveraged tools from stochastic geometry to model the distribution of users for a single, or multiple, time slots in wireless networks since the seminal paper of \cite{Baccelli1997}.
By considering the distribution of base stations (or users) as a point process, and computing the expectation of the corresponding functionals, closed form expressions for metrics
such as coverage and capacity can be obtained, a feat that could not be achieved through a purely information theoretic standpoint and has consequently lead to a proliferation in research on wireless networks.  For more discussions on how stochastic geometry is used to model wireless networks the reader is pointed towards the following \cite{HABDF09, Haenggi12, Baccelli2010, Schmidt2014, Kendall2010}  which offer a deeper insight. 

\subsection{Poisson point process}\label{s:ppp}
Prior to the use of stochastic geometry it was conventional to posit a hexagonal lattice to model the cells of macro base stations \cite{Ring1947}.
However, due to physical and economic constraints, the actual locations of base stations appear much more random.
The most commonly used point process is the  Poisson Point Process (PPP).  A simple PPP (there is at most one node at a single point in space) has been shown to represent the distribution of base stations in dense urban environments reasonably well \cite{Lu2015}.  

A PPP~\cite{LP17} $\Phi$ with intensity measure $\Lambda$ is defined by two properties:
\begin{description}
\item[(a)] The number of points in a set $A\subset X$, $\#(\Phi\cap A)$ is Poisson distributed with mean $\Lambda(A)$, or almost surely infinite if $\Lambda(A)=\infty$.
\item[(b)] If $\{A_i\}$ is a finite collection of disjoint regions, $\#(\Phi\cap A_i)$ are independent random variables.
\end{description}
Thus, the probability of a set $A$ containing $n$ points is
\begin{equation}
\mathbb{P}\left[\#(\Phi\cap A)=n\right]=\frac{\Lambda(A)^ne^{-\Lambda(A)}}{n!}
\end{equation}
For a uniform PPP, $\Lambda(A)$ is the density $\lambda$ multiplied by the $d$-dimensional volume of $A$.  For non-uniform PPP, $\Lambda$ can be a more general $\sigma$-finite non-atomic measure.  In either case, the mean number of nodes is $\bar{N}=\Lambda(X)$, which may be infinite.

One of the main tools used in the analysis is the Campbell-Mecke formula, 
\begin{align}\label{e:CM}
&\mathbb{E}_\Phi\left[\sum_{u \in \Phi} f(u) \prod_{v \in \Phi}g(v)\right]\\
&= \exp\left(-\int_{\mathbb{R}^n}(1 - g(x))\Lambda(\dd x)\right)
\int_{\mathbb{R}^n}f(x)g(x)\Lambda(\dd x)\nonumber
\end{align}
which reduces to either the probability generating functionals for PPPs (where there is no sum on the left hand side) and the widely used Campbell's theorem \cite{Haenggi12,baccelli2009} (where there is no product); an analysis of a more general functional was provided in \cite{Schilcher2016}.
Depending on the complexity of the model being studied determines whether the corresponding functionals can be given in closed form, but generally speaking analytic
expressions can be obtained when the distribution is assumed to be Poisson with uniform intensity measure. 

Another important property is that when points are randomly and independently thinned with probability $\wp$ the resultant process is Poisson with density $\wp \lambda$.
This models a simple channel access scheme where a user can transmit at each time slot with  probability $\wp$, and the result extends naturally to
location dependent thinning. 

As well as its tractability, the PPP is useful for constructing more complicated (interesting) point processes, see Cluster processes as an example.  Naturally, the deployment of base stations is not  spatially random since it is unlikely for two base stations to be built arbitrarily close together as interference effects will begin to dominate.  As such a typical network exhibits some sort of repulsion between points. 
Examples of random point processes that exhibit this type of behaviour include the Ginibre, Cox, Hard-Core and Gibbs processes which we summarise below.  

\subsection{Binomial point process}
A Binomial Point Process (BPP) can be obtained from a usual PPP by conditioning on the number of points in $\Phi$, which results in a loss in complete spatial randomness. 
However, both the void probability and  nearest neighbour distribution \cite{Andrews2011} (often needed when assuming a nearest neighbour association scheme) have a
simple analytic form (neglecting inhomogeneities and boundaries).  Results for the BPP are very similar to those for PPP when the number of points is large, and can be
obtained rigorously by ``De-Poissonization''~\cite{Penrose16}.  A BPP thinned as above is close to a PPP with the relevant intensity measure.

The BPP is easy to simulate: Choose $N$ points with respect to the (often uniform) normalised probability measure $\Lambda/\bar{N}$.  Thus to simulate a PPP, first choose
the total number of points as $N\sim \mbox{Poi}(\bar{N})$ and then a BPP with number of points $N$.  If the original measure is infinite, leading to $\bar{N}=\infty$ the
system must first be truncated to a finite region of interest.

\subsection{Gibbs point process}
A Gibbs point process is able to model the repulsion or regularity found within a network by having a density function that is defined by the pairwise interaction of points, and as such has various other applications such as modelling forestry statistics \cite{Stoyan2000}.
In a finite network of $n$ points the density function for a GPP is $f(\bx) = C \exp\left(-\sum_{i=1}^n \sum_{j, i<j} \phi(|\bx_i -\bx_j|)\right)$, where $C$ is a normalisation constant and $\phi$ is the \textit{pair potential function}.
One of the simpler examples from this family of point processes is the Strauss point process where $\phi = \mathbbm{1}_{r<r_0} \gamma$ where $\gamma \ge 0$.
Notice that for $\gamma = 0$ the model reduces to a PPP whilst when $\gamma = \infty$ it is a hard-core process; intuitively the parameter $\gamma$ determines the amount of spatial randomness in the model.  
Gibbs Point Processes \cite{Dereudre2017,Taylor2012,Guo2013}, which are closely related to the Gibbs statistical ensemble, have been shown to represent the SIR statistics and the Voronoi cell area distribution better than that of the typical Poisson model \cite{Taylor2012}; however they are significantly less tractable \cite{Guo2013}.

\subsection{Determinantal point process}

Determinantal point processes (DPPs) have recently been proposed to better model the intrinsic intrinsic repulsion exhibited in the distribution of base stations \cite{HABDF09, goldman2010palm, Shirai2013, Deng2015} whilst still retaining some tractability.
DPPs were first introduced in \cite{macchi1975coincidence, macchi1977fermion} to study the distribution of fermions in thermal equilibrium \cite{decreusefond2016determinantal},  as a consequence DPPs were originally referred to as fermion point processes, but also arise in the eigenvalues of random matrices,  quantum mechanics, representation theory, spanning trees, self-avoiding random walks \cite{soshnikov2000determinantal, torquato2008point, hough2006determinantal} and more recently have been used in machine learning\cite{Kulesza2012}. 
Generally speaking a DPP is defined by the $n-$th order product densities and naturally exhibit repulsion between points creating a more regular structure; for the exact details of the DPP the reader is referred to \cite{soshnikov2000determinantal,  hough2006determinantal}.
As an aside, the permanental process (which the Cox process is an example) is the natural counterpart to the the DPP where points tend to cluster together and have been used in the study of bosons \cite{hough2006determinantal}.

A Ginibre point process (GPPs) is a DPP on $\mathbb{C}$, first introduced in Ref \cite{ginibre1965statistical}, which is an extension of the Dyson point process (a DPP on $\mathbb{R}$).
The GPP is characterised by the $n^{\text{th}}$ order product densities, defined on $\mathbb{C}$, given by, 
\es{\rho^{(n)}(x_1, ...,x_n) = \text{det}(K(x_i,x_j))_{1\le i,j\le n}\:\:}
where  $(a_{ij})_{1\le i,j\le n}$ is the usual matrix notation,  det is the determinant of that matrix and $K$ is the Gaussian kernel 
\es{K(x,y) = \frac{1}{\pi}e^{x\bar{y}}e^{-\frac{|x|^2+|y|^2}{2}}\:\:\: x, y \in \mathbb{C}} 
with respect to the Lebesgue measure on $\mathbb{C}$ \cite{blaszczyszyn2018stochastic, Schmidt2014}. 
The first order correlation function is simply $\rho^{(1)}(x) = \pi^{-1}$, which is the usual density,  the second moment density simplifies due to motion invariance and only depends on the distance  between pairs of  points, $\rho^{(2)}(x, y) = \rho^{(2)}(|x-y|)$ whilst it is also straight forward to show the repulsive nature of the point process since 
the pair correlation function is strictly less than one \cite{blaszczyszyn2018stochastic, Deng2015}. 

 A thinning of the normal Ginibre point process (see \cite{Lavancier2015}) has also been studied  to allow for interpolation between the original GPP ($\beta = 1$) and the PPP which it converges to weakly as $\beta \to 0$; the parameter $\beta$ can be seen to determine the level of \textit{repulsiveness} exhibited by the base stations. \\

In \cite{Deng2015} the authors showed for a $\beta$-Ginibre process, by adjusting  $\beta$ to best fit the distribution of Base Stations the derived integral representation of the coverage probability is a better fit compared with the typical Poisson model. 
Similar results were obtained for the wider class of DPPs, of which Ginibre is an example \cite{Li2015}. 

\subsection{Mat\'ern point process}
A hard-core process creates a more regular distribution of points. 
One such example is the Mat\'ern Type I \cite{matern2013spatial} where points are deleted from a uniform PPP if it has a neighbour within some distance $r_0$; the transmission scheme is a spatially dependent thinning.  
This type of process ensures two transmitters aren't too close to one another and transmitting.

\subsection{Cox point process}
Many networks can often exhibit some form of clustering; for example in more urban environments where people tend to gravitate around popular places such as shopping centres and sporting events \cite{Ying2014}.
In 5G networks, which are likely to be extremely dense in order to deliver the desired throughput, it is probable that within a city there will be clustering of smaller access points (femto or pico cells) around places of work and retail areas, one possible model for this is a Cox process.
A Cox (or doubly Stochastic) process  can be viewed as a PPP with random intensity measure\cite{grandell2006doubly, moller2006properties}.
The appearance of a point in space is likely the result of a large intensity measure, the local neighbourhood is likely to also have a large intensity measure and thus be populated by many other nodes which form a cluster\cite{blaszczyszyn2018stochastic}.
By conditioning on the random driving measure the Cox process reduces to a PPP; whilst a Cox process can be obtained from a PPP by applying a random thinning (distinct from the deterministic  thinning mentioned earlier) \cite{moller2006properties,Haenggi12}.

An interesting application of these doubly stochastic processes is to vehicular networks.
For example, in Ref\cite{chetlur2018coverage} they model the random locations of vehicles by a 1D process on a road, with each road being modelled by a line process, allowing for a more realistic, but still tractable, model for connectivity in vehicular networks.

\subsection{Cluster point processes}
Another family of point processes that exhibits clustering, and has some overlap with the cox process, are cluster point processes. 
In general a cluster process can be formed by first generating a parent process $\Phi_p$, and for each point in that process generate another point process independent from each other (daughter PP), then the cluster point process is  the union of all daughter process generated from $\Phi_p$; a Poisson cluster process is a special type of this more general cluster process.
Mathematically speaking, if $\Phi_p$ is the parent point process with $n$ points, with locations $\{x_1,x_2,...\}$ and $\Phi_i$ is the family of finite daughter processes corresponding to each parent (untranslated), then the cluster process is the resulting union, $\Phi = \cup_{i \in [n]} \Phi_i + x_i$ \cite{Haenggi12}.

One of the  simpler cluster models is the Neyman-Scott process where the parent points follow a PPP in $\mathcal{A}$ and the daughter nodes are distributed according to a uniform PPP in $\mathcal{B} \subset \mathcal{A}$. 
Some work has been done on these processes in a wireless network context with interference but often result in complex expressions for metrics such as the mean achievable rate and success probability  \cite{Ganti2009, Zhong2013}.
In summary, there are many interesting and applicable generalisations of the PPP, but as usual there is a trade-off between realism and tractability.

\section{Random links}
\subsection{Geometry}
Having described possible point processes that define the locations of the nodes, we now discuss the links, and in particular, the statement that in spatial networks, links occur between closely spaced nodes.  For this, we need a notion of distance, the map ${\cal D}:X\times X\to \mathbb{R}$.
The distance function ${\cal D}$ is normally Euclidean distance, or for a flat torus, the shortest Euclidean distance taking account of the identification of the opposite sides/faces.  More general metric spaces are natural, but note that each of the metric space axioms is violated in natural examples arising in wireless communications: (a) ${\cal D}(x,y)=0$ for non-equal $x$ and $y$ if the latter correspond to the same position but different orientations for the case of anisotropic networks; see Sec.~\ref{s:ani}. (b) The triangle inequality is violated if we do not allow transmission through obstacles, i.e. such a path has infinite effective distance; see Sec.~\ref{s:nc}.  (c) ${\cal D}(x,y)\neq{\cal D}(y,x)$ if, due to differing transmission power, we can transmit a signal from $x$ to $y$ but not vice versa; this leads naturally to directed graphs, but appears relatively unexplored in the context of RGG; see for example Ref.~\cite{LDWGZW09,Shang10,FBG14}.

Motivated by real data, social networks, and in an attempt to understand the effects of geometry on spatial network properties, recently works have extended RGG ideas from flat to hyperbolic spaces \cite{Gugelmann2012}.
Other works have  suggested that the distribution of the network link-lengths may be characterised by dimensions higher than that of the embedding space and hence there is a dimensionality reduction when reconstructing a network from a dataset \cite{Daqing2011}.
For wireless network applications, non-Euclidean geometries have also been proposed as an alternative virtual embedding that maps the network into a hyperbolic space that can support more efficient packet routing schemes \cite{Cvetkovski2009}.
With the exception of a few other similar works however, the majority of RGGs literature with applications to wireless networks is concerned with $d=2$ or $d=3$ dimensional Euclidean geometries in $\mathbb{R}^d$.

\subsection{Pair connection functions and soft random geometric graphs}\label{s:connection_fns}
A natural generalisation of RGG, both from a mathematical and a practical point of view, is that of soft random geometric graphs (SRGG).  Here, links between nodes are made independently, with a probability given by a function $H(r)$, the (pair) connection function, of the mutual distance $r$.  Closely related models are given different names by different communities in the literature: The first such model was the Waxman graph~\cite{Waxman88}, followed by continuum percolation~\cite{Penrose91,Alexander93}.  Recent literature has used the SRGG label~\cite{Penrose16,Krioukov16,MP15} (which we prefer as it is most specific), as well as random connection models~\cite{Bradonjic14,iyer2018random,Mao17} and spatially embedded random networks~\cite{BDB07,PR17,HA17}.

Some connection functions found in the literature are chosen for mathematical simplicity, such as the unit disk (RGG model) $H(r)=\mathbbm{1}_{r<r_0}$, constant
(Erd\H{o}s-R\'enyi model) $H(r)=p\in(0,1)$ or other piecewise linear functions.  Others are chosen from the physical characteristics of wireless communication channels.  
For example, we can write the link probability in a noisy environment with interference as, 
\es{H(r)=\mathbb{P}(\mbox{SINR}>q)}
where $q$ is a constant threshold and SINR is the signal to interference plus noise ratio, where the signal (and interfering signals) is proportional to the product of the transmitter and receiver gains ($G_T$ and $G_R$ respectively),
the (random) channel gain $|h|^2$ and the path loss $r^{-\eta}$.  Here $\eta=2$ corresponds to the (free space) inverse square law, but values typically in the range
$2\leq\eta \leq 6$ have been found empirically for cluttered environments.
More specifically, for a network where all nodes are trying to transmit data on the same channel, the probability a link can be formed at a particular instance between the receiver and intended transmitter $i$, with separation distance $r_i$, is given by,
\begin{align}\label{e:sinr}
H(r_i)&=\mathbb{P}\left(\frac{G_{T_i}G_R|h_i|^2 r_i^{-\eta}}{\mathcal{N} + \gamma \sum_{k \neq i}G_{T_k}G_R|h_k|^2r_k^{-\eta}}>q\right)
\end{align} 
The parameter $\gamma \in [0,1]$ is introduced to apply a random thinning of the interfering signals representing an ALOHA transmission scheme, which is the simplest example of how the set of interferers changes with time \cite{abramson1970aloha}. 
In the ALOHA model devices are active with probability $\gamma \in [0,1]$ with $\gamma (1-\gamma)$ being the probability that a device is on (off) or, when devices have a single antenna, transmitting (receiving). 
Therefore, the worst case scenario, being $\gamma = 1$, is where all users in the network try to concurrently transmit on the same frequency.

Firstly, since the interference is measured at the receiver the network can become highly directional with links occurring in one direction but not the other \cite{GWBDC15}. 
In addition, Eq.\ref{e:sinr} is not strictly speaking a pair connection function as it depends on the locations of all nodes in the point process, and the location of the receiver, and not just the receiver-transmitter distance.
That said, by taking the spatial average of the set of interfering nodes through equation \eqref{e:CM} it can be expressed as a function of receiver and transmitter position, which, for point processes that are translational invariant, can be written as the standard pair-connection function introduced earlier . 

Under the Rayleigh fading assumption (diffuse propagation), the channel gain is
exponentially distributed, and assuming there is no interference within the network ($\gamma = 0$),  leads to 

\begin{equation}
H(r)=e^{-(r/r_0)^\eta}
\end{equation}
  We note that $\eta=1$ is the Waxman~\cite{Waxman88} connection function and $\eta\to\infty$ is the RGG. \Ck{}

There are many more complicated channel models,
including Rician fading (a combination of specular and diffuse propagation) and MIMO (multi-antenna devices).  Ref.~\cite{DG16} tabulates many of these mathematical and physical connection functions. Ref.~\cite{DG17} shows how to extend these short ranged connection functions to a ``small world'' model with random long ranged links  which models networks that have both spatial and non-spatial links. For example, the non-spatial links may represent a limited infrastructure which can appear non-spatial over such large distances, which have the added benefit of reducing the average hop count \cite{sharma2005hybrid}. 
Finally, connection functions can be constructed empirically with any spatial network for which the links can be assumed to be random~\cite{WDKD16}.

\subsection{Effects of differing connection functions}
Qualitatively, many connection functions look alike, for example Rayleigh is often a good approximation for Rician~\cite{BDC13}.  Furthermore, the connection probability can be well approximated
by a formula involving the connection function only via two of its moments~\cite{DG16}.  These results suggest that SRGG properties are only mildly dependent on the form of the connection function.  There are however several qualitative effects that depend on the connection function:

The (hard) RGG differs from SRGG in several qualitative ways. At extremely high density, two isolated nodes are likely to be close in a SRGG, but they cannot be in an RGG.  In order to have two nodes separated from the rest (and either both isolated or mutually linked), an RGG requires only a small extra region free of other nodes, whilst an SRGG requires the absence of (at high density) many extra links.  This means that the corrections to the first order (single isolated node) approximation for the connection probability are algebraic for the RGG but exponential (i.e., much smaller) for the SRGG~\cite{CDG12b}.  
There are also numerical and qualitative results showing that the k-connectivity (a graph is said to be k-connected if there exist k mutually independent paths between any two nodes in the network) is much better approximated by the minimum degree in SRGG than RGG~\cite{GDC14}. 
Last but not least, the presence of a second source of randomness, the links, in SRGG permits the study of metadistributions and entropy conditional on the node locations; see Secs.~\ref{s:meta} and~\ref{s:entropy} .

For the Rayleigh connection function, there are a number of qualitative transitions when the path loss exponent $\eta$ becomes equal to the spatial dimension $d$.  Directional radiation
patterns are modelled by giving nodes orientations as well as locations, and making the gains $G_T$ and/or $G_R$ depend on the orientations relative to the line joining the two nodes.
They lead to higher connectivity than the isotropic pattern given fixed total power if $\eta<d$~\cite{CD13,GDC14b}.  Interference from distant nodes diverges (Olbers' paradox)
if $\eta\leq d$~\cite{Shepard96}; of course practical systems have a finite number of nodes, however this means that the properties related to the interference depend measurably
on the overall size and shape of the domain.    The quantity $d/\eta$ also appears as a scaling exponent of the connectivity with respect to power or number of antennas~\cite{CGD15}. 

For connection functions with algebraic decay, $H(r)\sim r^{-\alpha}$ as $r\to\infty$ the critical value is again the spatial dimension.  If $\alpha\leq d$ the mean degree
in an infinite space  $\int_0^\infty H(r) S_d r^{d-1}dr$ diverges, and thus all graph properties in finite domains depend sensitively on the size and shape of the domain.
For the SRGG in one dimension, the value $\alpha=2$ is also critical, as for
$\alpha\leq 2$ it is possible for the system to percolate, that is, contain an infinite cluster, despite the presence of arbitrarily large gaps.  The literature here is for closely related
lattice models~\cite{NS86}.

\section{Connectivity}

\subsection{Scaling limits}
A RGG (or SRGG) typically has three characteristic length scales, the system size $L$ (if finite), the connection range $r_0$ (SRGG with long range connection function may
be an exception), and the typical distance between nodes $\lambda^{-1/d}$ where $\lambda$ is the mean density, equal to  $\Lambda(\mathcal{V})/|\mathcal{V}|$ if the latter is defined.
If we scale the system, all of these lengths scale by the same factor, so clearly only the ratios are relevant; all relevant features may be represented in terms of the dimensionless
parameters $\lambda r_0^d$ (proportional to the mean degree) and $\lambda L^d$ (proportional to the total number of nodes).   

It is difficult to make quantitative rigorous statements about finite RGG, that is, where the average total number of nodes $\bar{N}=\Lambda(\mathcal{V})$ is finite.  So, most
rigorous statements, for example, reviewed in Ref.~\cite{Walters11}, are for infinite RGG, for example on $\mathbb{R}^2$, or on a sequence of RGG in finite domains for which $\bar{N}\to\infty$.  On a cube $[0,L]^d$,
we have $|\mathcal{V}|=L^d$, so a diverging number of nodes corresponds to $\lambda L^d\to\infty$.  The remaining dimensionless parameter $\lambda r_0^d$ may either remain
constant, decrease, or, more commonly, also diverge at an arbitrary rate with respect to $\lambda L^d$. In the literature, either $r_0$ or $L$ is usually held fixed, with the
remaining two quantities allowed to vary, to obtain the above limits.  If $r_0$ varies, the connection function takes the form $H(r)=h(r/r_0)$.
We now discuss some key results on percolation and full connectivity under different scaling regimes. 

\subsection{Percolation}\label{s:perc}
In infinite systems, including the first paper, Ref \cite{Gilbert61} on RGGs in $\R^2$, there are an infinite number of nodes, so the only parameter is $\lambda r_0^d$.
Here, we will follow the usual convention and set $r_0=1$.  
As $\lambda$ increases, there is a transition, known as percolation~\cite{Grimmett99}, from a state in which all the connected components are almost surely finite, to one in which
there is almost surely an infinite component.  Percolation was already known in the context of lattices~\cite{BH57}, so the RGG was an early example of ``continuum percolation.''
Many results in RGG percolation are obtained using the more well-studied lattice problems.

An initial lower bound on the critical parameter was provided by relating the PPP to a branching process, whilst an initial upper bound was achieved by tiling the plane and drawing
from results on bond percolation on the square lattice. 
In fact Gilbert also outlines that through tessellation of the plane with a hexagonal lattice the upper bound could be improved upon, although this was conditional on the critical probability for the triangular lattice being $1/2$ (which was later shown by \cite{kesten1982percolation}). 
A more detailed discussion of percolation in RGGs can be found in \cite{Walters11}.
A natural extension is to SRGGs.

Surprisingly, percolation on SRGGs has only recently been studied for the perspective of networks. 
Instead, motivated by physical and chemical models of composite materials, micro-emulsions, and liquids, much work has been done on spherical particle interactions, namely those with a hard core and a soft shell. 
These approaches would either attempt to map the system to lattice percolation, where results are well known, or would resort to extensive numerical Monte Carlo simulations \cite{isichenko1992percolation}.
The particle hard core was said to be an impenetrable portion  while the soft shell was associated with the range across which an particle interaction is allowed, (e.g., a charge transfer or excitation) \cite{bug1985interactions}. 
Percolation thresholds therefore apply in this setting and can potentially be linked to observable material properties \cite{balberg1987invariant}. 

An interesting extension is to ask the same percolation question but on an SINR graph, thus incorporating interference;  nodes are deployed according to some point process and links formed
 according to the SINR connection function, Eq.~\ref{e:sinr}. Initial investigations have drawn inspiration from continuum percolation with interference to derive for example the
network capacity \cite{FDTT07}. 
This model of continuum percolation in an SINR graph has the additional complexity of the link probabilities between any two nodes depending on the underlying Point Process of nodes and the assumed connection model.
First note that we should expect percolation in the RGG with soft connectivity (as for the case with hard connectivity) to be monotonic in $\lambda$, such that the probability of percolation is zero when $\lambda < \lambda_c$, and the $\lambda > \lambda_c$ the graph percolates almost surely. 
An SINR graph can be shown to percolate for a small enough $\gamma>0$ provided $\lambda \ge \lambda_c$ and the path loss function is integrable on $\R^2\backslash B_0(\epsilon)$ \cite{Dousse2006}.
Namely, by randomly thinning the number of nodes transmitting concurrently on the same channel the graph can percolate, provided the density is larger than the critical density needed for percolation when there is no interference.
In a similar vein, \cite{Vaze12} showed that for small enough $q$ (recall $q$ is the minimum threshold value for the signal measured at the receiver in order for there to be a successful link, see Sec\ref{s:connection_fns}), assuming a non-singular path-loss with no thinning ($\gamma = 1$), percolation only occurs for a certain interval of densities; when the number of nodes in the network is small  percolation is obstructed due to large gaps, whilst when the density is too large interference effects begin to dominate.
So once again, it is possible for the network to transition from sub-critical to super-critical back to sub-critical again as the node density increases for a fixed set of network parameters.
However, we should expect this behaviour to break down for the singular path loss model, that is to say once percolation is achieved adding more nodes (therefore interferes) has no bearing; the mean degree is monotonic in $\lambda$ \cite{Vaze12,Pratt16}.

\subsection{Isolation and connectivity}
The probability that a node in a RGG or SRGG at location $\bf r$ is isolated, that is, has no links, is easily found from Eq.~(\ref{e:CM}) to be
\es{
P_{iso}({\bf r})=e^{-\lambda M({\bf r})}
}
with 
\es{
M(\mathbf{r}) = \int_\mathcal{V} h(|\mathbf{r}-\mathbf{r}'|/r_0) \text{d}\mathbf{r}'
}
which is often called the position dependent connectivity mass, and $\lambda M$ the expected degree at $\mathbf{r}$.
If $M({\bf r})$ is finite, true except for very long ranged connection functions in infinite domains,
this probability is non-zero.  The expected number of isolated nodes is thus
\es{
\mathbb{E}(N_1)=\lambda \int_{\cal V} P_{iso}({\bf r})d{\bf r}}
denoting the number of components of size $m$ by $N_m$.  For non-uniform Poisson processes, these equations hold with $\lambda d{\bf r}$ replaced by integration
over the more general measure $d\Lambda({\bf r})$.

Here, we are concerned with the probability that the network is connected $\mathbb{P}(N_N=1)$, that is, there exists a multihop path between any pair of nodes.
In much of the literature this property is called ``fully connected' to avoid confusion with ``connected'' used to indicate individual links between nodes, but here we keep
to standard terminology in graph/network theory and denote the overall property as connectivity.  

By looking at isolated nodes, we see under very mild assumptions that isolation of distant nodes is effectively independent, and hence that if there are infinitely many nodes
and the integral is finite (connection function not too long ranged), the probability that the network is connected , that is, there exists a multihop path between any pair of
nodes, is zero.  Thus, connectivity is normally considered only for finite networks.  In terms of the length scales above, we see that
$M\approx C r_0^d$, where the constant $C$ is an integral of the function $h$.  Thus the number of isolated nodes
is roughly $\lambda L^d \exp\left(-C\lambda r_0^d\right)$, showing that a scaling in which the mean degree is roughly the logarithm of the
number of nodes leads to the number of isolated nodes of order unity.  A more precise argument to fix the constant needs to take boundary effects
into account; see the next subsection.

Isolated nodes provide not only a bound but in fact the key to understanding connectivity for $d\geq 2$: It turns out that they are the main obstruction to connectivity.
More precisely, in this connectivity scaling limit, the probability of clusters other than isolated nodes or the remaining large component is negligible, so we have the
connection probability
\es{
\mathbb{P}(N_N=1) \approx \exp \pr{-\lambda \int_{\mathcal{V}} e^{-\lambda M(\mathbf{r})} \text{d}\mathbf{r}}
\label{pfc}}
This formula has been proved exactly in the connectivity scaling limit for RGG and in some cases for SRGG \cite{Penrose16}; see also Refs.~\cite{MA13,iyer2018random}. Numerical
simulations confirm its validity well beyond what can be proven. It also holds for the binomial point process in the same limit (where $\lambda$
is now $N/|{\cal V}|$).  An analogous result is found in random graphs with no spatial component \cite{Bollobas1998,Erdos1960}.  Some literature approximates
the outer exponential for very high connection probability, that is, $\exp(-z)\approx 1-z$.  We discuss $d=1$ briefly in Sec \ref{s:1d} below.

For comparison, the percolation transition occurs when the mean degree is a constant, whilst the connectivity transition occurs at significantly higher
densities, when it is logarithmic in the total number of nodes.  For a discussion on connectivity in different limits see Refs. \cite{MA12, DG16}. 

\subsection{Connectivity and boundaries}\label{s:boundaries}
The dominant contribution to the integral in Eq.~\ref{pfc} comes from regions of small connectivity mass, that is, near the boundaries, especially corners.  Intuitively, nodes near the corners have fewer neighbours on average and are more likely to be isolated.  This is however balanced by the fact that there are fewer nodes near the corners.  In the connectivity
scaling limit, the dominant contribution comes from the bulk in two dimensions (though edges for the related problem of $k$-connectivity) and from two dimensional faces in three dimensions; see Refs.~\cite{Penrose03,Walters11}.  Thus, in two dimensions (particularly) there has been some justification for neglecting boundary effects, ignoring nodes
near the boundary \cite{Bettstetter2005a}.  In another example \cite{Andrews2011}, the authors  analysed different cellular network metrics assuming a uniform PPP deployment
of base stations on the plane. Their findings concentrate on the coverage experienced by the \textit{typical user}, a concept that follows from the translational invariance of a
uniform PPP.

However, realistic networks have a finite number of nodes, and except for spherical and (physically questionable) toroidal domains, some of these lie close to a boundary.
The connectivity scaling gives an exponential system size $L$ as a function of density $\lambda$ at fixed $r_0$.
Thus in realistic networks, effects of different types of boundary need to be considered.
Using a spatial decomposition argument, the connection probability was considered using a sum of contributions from different boundary elements: a bulk component, and
edge and corner contributions that depend on the node density and the connection function $H(r)$ \cite{CDG12b}. 
Keeping only the leading order contributions for each boundary component \cite{CDG12a} we can obtain expressions for a variety of $d\geq 2$ dimensional domains, e.g., a
3D prism in the shape of a monopoly house \cite{CGD14}
\es{
\mathbb{P}(N_N=1)= \exp \prr{-\sum_{i} \sum_{b\in \mathcal{B}_i}  \lambda^{1-i} G_{d,i}^{(b)}  V_b  e^{-\lambda \Omega_b H_{d-1}}}
\label{pfc2}}
where $0\leq i \leq d$ is the co-dimension of a boundary component $b$ (e.g, an edge or a surface); $B_i$ is the set of boundaries with co-dimension $i$; $G_{d,i}^{(b)}$ is a geometrical factor obtained by expanding equation \eqref{pfc} in the vicinity of the boundary component $b$; $V_b$ is the $(d-i)$-dimensional volume of the component $b$; $\Omega_b$ is the magnitude of the available angular region of a boundary component $b$ (i.e., the solid angle it subtends); and $H_{d-1}$ is the $(d-1)$th moment of $H(r)$.  Ref.~\cite{DG16} tightened a number of the arguments and showed
that the geometrical factor could also be written as another moment $H_{d-2}$ of the connection function $H(r)$, that is, the connection probability depends on the connection
function only through two of its moments. Curvature effects, i.e., when the boundary components are not straight or flat, offer corrective, second order, contributions to each
boundary component \cite{DG16}; an interesting exception are cusps (zero-angle corners) which behave like one dimensional systems (see below).
The derivation of the coefficients of each boundary component is slightly more lengthy, however one can follow a structured approach \cite{DG16},
for both k-connectivity \cite{Khalid2014} and three dimensional domains \cite{Khalid2013}. 

In interference limited networks, boundaries cause a reduction in the interference field thereby improving coverage near border users \cite{GWBDC15, CDG12a}.
In contrast, if users are mobile e.g. follow a random waypoint model (see Sec.~\ref{s:rw}) border users are more likely to be in outage \cite{Pratt16}.

\subsection{Connectivity in one dimension}\label{s:1d}
In one dimension the isolated nodes are less relevant in the high density limit and the network may instead break full connectivity by splitting
into two or more large pieces. For the RGG unit disk model it is relatively straightforward to calculate the probability of a gap of given size~\cite{Devroye81,MH13}, but for soft connection functions it remains an interesting open  problem. One dimensional effects may appear in higher dimensional systems, notably annuli or rectangles with very large ratio of
length to width, and also cusps, that is, curved boundaries meeting at a corner with vanishing angle.

\subsection{Applications of connectivity}
The connection probability is often associated with wireless network reliability, e.g., to prevent disruption due to short radio range, wireless 
node sparsity, energy resources, cyber attacks, random failures, background noise, etc.
The analytical results described above are needed if one is interested for example in analysing the reliability assurance of wireless networks \cite{Laranjeira2014} or when developing wireless security and trust protocols that are tailored to specific network deployments \cite{Coon2014,koufos2019boundaries}, or when considering routing protocols in wireless sensor networks \cite{Fu2017}.
Also,global and local  connectivity metrics can be used to enhance wireless localization when positioning devices like GPS are unavailable \cite{Nguyen2015}.
Typically for instance, after an initialization phase, wireless nodes know with whom they can directly communicate, but have no idea about their relative geographic locations within the deployment region.
There is however enough geometry information encoded in the connectivity structure of the network to identify topological features like boundaries \cite{Funke2006}.
With some modifications in $H(r)$, the above methods can also be used to estimate epidemic spreading rates in SIR networks \cite{Estrada2016}.
Many other reliability metrics like $P(\text{path})$ defined as the percentage of nodes that are connected via a multi-hop fashion are equivalent to
the connection probability at high node densities but will not be discussed in detail here.

\section{Obstacles and Reflections}\label{s:nc}

The mathematical treatment of network connectivity has undergone quite a transformation when going from infinite domains to finite ones.
Namely, while going from equation \eqref{pfc} to \eqref{pfc2}, interesting topological features of the network were uncovered and directly related to boundary element components in a mathematically tractable manner.
Effectively, these can be described as geometrical constraints to the wireless network deployment region that in turn affect the wireless connectivity and performance of the network. 
Another closely related boundary effect is that caused by blockage, such as buildings obstructing the direct line of sight (LoS) propagation of radio and higher frequency signals.
Therefore, in this subsection we will discuss the extension of the above efforts towards non-convex deployment regions.

There are three popular approaches to incorporate blockage effects to the modelling of wireless networks. 
One method is using ray tracing to perform site-specific simulations \cite{Schaubach1992}. 
This however requires a lot of accurate site information, such as the size and location of blockages in order to generate the received wireless signal strengths at each wireless device. 
Ray tracing techniques therefore trades the complexity of numerical computation for an accurate site-specific solution. 
In many instances this is necessary and achievable.
For example, with the advancement of current imaging technologies and the availability of accurate datasets of digitized 3D maps ray tracing methods will eventually become commonplace.

The second approach is to establish some stochastic model, e.g., a random variable modelling the statistical characteristics of blockage effects on wireless propagation. 
An advantage of such an approach is that it can be easily incorporated into existing stochastic geometry models \cite{Baccelli2010} and can therefore be used to analyse general networks. 
For example urban areas can be modelled as random lattices \cite{Marano1999} that are blocked with some probability, and may be assembled into random shapes whose blockage effects are then encapsulated as a random variable contribution towards the received signal to interference ratio between each network pair \cite{Bai2014}.

The third approach builds on the intuition afforded by the previous subsection and equation \eqref{pfc2}, 
namely, to evoke a spatial decomposition argument and partition the network deployment domain $\mathcal{V}$ into separate connectivity contributions.
For example if the domain is naturally split into two or more well separated domains joint only via small keyhole like openings, then one can be concerned with the connectivity metrics, e.g.,  $P_{fc}$ in each sub-domain, linked together by some inter-domain connectivity function \cite{Georgiou2013}. 
The latter, similar to the case of convex domains, depends on the size and shape of the opening, the node density, and the connection function $H(r)$.

Crucial to the refinement needed to expand the above framework towards non-convex domains is that the LoS constraint can be realized through the introduction of a characteristic function $\chi(\mathbf{r},\mathbf{r}')$ that equals one if a non-obstructed line of sight exists between points $\mathbf{r}$ and $\mathbf{r}'$, and zero otherwise. 
For example, the connectivity mass of a node located at $\mathbf{r}$ can be modified as follows
\es{
M(\mathbf{r}) = \int_{\mathcal{V}} \chi(\mathbf{r},\mathbf{r}')H(|\mathbf{r}-\mathbf{r}'|) \text{d}\mathbf{r}'
\label{M}}
and inserted into \eqref{pfc} \cite{Giles2016}. 
Note that non-convex domains have not been studied rigorously yet \cite{Penrose16}.
Expanding in a similar manner as in \cite{DG16} and keeping only leading order contributions for each boundary component (including the non-convex ones) the expression of \eqref{pfc2} remains valid.
For instance, Ref. \cite{Giles2016} studied the case of circular and spherical obstacles.  The effect of large obstacles is similar to that of curvature in the domain boundary; small
obstacles have a differing mathematical formulation and have a small effect on connectivity unless they are very numerous.

A further modification to equation \eqref{M} is that of including reflection effects. Here, pairs of nodes that would otherwise not connect due to a direct LoS path, are allowed to connect if a reflected LoS exists \cite{GBRDC15}. 

Keyhole domains involve large connected regions with a small gap through which a signal may pass; they require different techniques depending on the relative size of the various length scales (size of hole as well as connection range, typical distance between nodes and overall system size); see Ref.~\cite{Georgiou2013}.

\section{Directional Antennas}\label{s:ani}

Much of the work on network connectivity that can be found in the literature assumes that each wireless node, whether a small mobile device or a large base station,
radiates its power and therefore information isotropically, i.e., uniformly in all directions.  In two dimensions this can be achieved by an axially symmetric antenna
(though most antennas are not symmetric), but in three dimensions it is not physically possible to radiate electromagnetic waves uniformly.
Nevertheless, a theoretical isotropic antenna is often used as a reference antenna for measuring and characterising the antenna gain $G(\theta)$ of real devices and is
specified in dBi, or decibels over isotropic. 
This refers to the power in some direction $\theta$ divided by the power that would have been transmitted by an isotropic antenna emitting at the same total power.

Anisotropic radiation gain profiles allow wireless communication links to be established along longer distances in their boresight (strongest) direction.
A trade-off usually exists however in other directions which have a lower antenna gain.
To see this, let us assume negligible inter-node interference, and define the connection probability between transmitting node $i$ and receiving node $j$ through the relation
$H_{ij}=\mathbb{P}(\textrm{SNR} \cdot |h|^2 \geq \wp)$,
where SNR denotes the long-term average received signal-to-noise ratio and $h$ is the channel transfer coefficient for single input single output (SISO) antenna systems. 
 
Assuming lossless antennas, then the signal power at the receiver, given by the Friis transmission formula  \cite{balanis1992antenna}, gives 
$
\textrm{SNR} \propto G_{ij} G_{ji}  r_{ij}^{-\eta}
$,
where $r_{ij}=|\br_i - \br_j|$ is the Euclidean distance between the two nodes, $\eta$ is the path loss exponent (typically $\eta\geq2$), and
$G_{ij}$ is the gain of the antenna $i$ observed in the direction of node $j$.
Isotropic radiation patterns have a constant gain $G=1$, while anisotropic ones are functions of the polar angle $\theta$, appropriately normalized by the condition $\int_{0}^{2\pi } G(\theta) \dd \theta =2\pi$ in two dimensions and $\int_0^{2\pi} \int_0^\pi  G(\theta,\phi) \sin \theta \dd\theta \dd \phi=4\pi$ in three.
Now, as a simple example we may approximate a microstrip (patch) antenna gain profile by a cardioid function in two dimensions by \cite{balanis2011modern}
\es{
G_{ij}( \theta_{ij})=1 +\epsilon \cos \theta_{ij},
\label{cc}}
where $\epsilon\in[0,1]$ measures the extent of deformation from the isotropic case ($\epsilon=0$), and $\theta_{ij}$ is the direction of receiving node $j$ relative to the antenna orientation of node $i$. 
It follows that the connection function is now explicitly and quite strongly dependent on the angles and orientations of the wireless nodes.
Note that if the channel gain $|h|^2$ is assumed to be exponentially distributed as to model for example a Rayleigh fading channel, the connection function is then also exponentially dependent on the antenna gain parameters $\theta$ and $\epsilon$.
In a similar way, one can define other smooth functions that approximate the gain profiles of various other antennas (e.g., horn and dipole) as well as multi-directional, e.g., by modifying $G= 1+\epsilon \cos n \theta$ with $n>1$ equally spaced and identical lobes.
An alternative approach to smooth gain profiles is to use sectorised models \cite{Dai2013} or keyhole models \cite{Li2011}.
Both models are somewhat over-simplified and may not be able to capture in full for example the nulling capability of realistic antennas or may ignore any side or back lobes \cite{Wang2017}.
Regardless, it is still possible to study in some depth and to some accuracy various wireless network properties such as their capacity \cite{Yi2003}, power consumption \cite{Nasipuri2002}, security \cite{Hu2004}, and medium access control (MAC) protocol (MAC) design and efficiency \cite{Singh2011}.

The impact of directional antennas on the connectivity of the resulting network topology encapsulates all the above wireless network properties. 
It is well accepted that a better connected network will have shorter multihop paths from source to destination, use less energy, will access the wireless medium less often, cause less interference, and have higher throughput.
This is however an idealized scenario in what is often a very large and complex multi-layered network system.
Therefore, the impact of directional antennas on the connectivity of wireless networks has been studied both analytically and numerically in both ad hoc and cellular networks and has unveiled a number of interesting results.

Since ad hoc networks operate in a decentralized and self-organizing manner, it can be assumed that in most such cases antenna orientations are either random from deployment (e.g. air-dropped sensor devices) or are randomly chosen from a set of possible configurations without any coordination with other nodes. 
In such instances, randomized and greedy beamforming approaches improve ad hoc network connectivity under certain circumstances
\cite{Bettstetter2005a,Koskinen2006}.
Namely, it was shown that while directional antennas and beamforming can significantly improve point-to-point wireless links when perfectly aligned, when antenna orientations are chosen at random the 1-hop network connectivity is typically  deteriorated when the path loss exponent $\eta>d$ in $d=2,3,$ dimensions \cite{Coon2013}.  
In contrast, multihop connectivity is greatly improved,  especially in the dense regime as can be seen by the decreasing number of hops (relays) needed between stations \cite{Georgiou2015b}. 
Implicit for this return, is perfect interference management and a good MAC protocol ,which controls the user access to the channel \cite{Ko2000}. 
Moving to higher dimensions and confined or indoor geometries have also been studied from a network connectivity perspective \cite{Georgiou2014}.
Interestingly, boundary effects can significantly deteriorate coverage if antenna orientation is randomly chosen. 
In contrast, when considering interference effects in ad hoc networks, directional antennas double up in benefits since 
transmitters cause less interference while having a longer reach, and also receivers can null out interfering signals from unwanted directions \cite{Georgiou2015c} . 
These kind of insights and statistics are useful when designing wireless sensor networks (WSNs) and wanting to choose the right density of directional nodes to be deployed in order to meet certain connectivity requirements.

Advancements in beamforming and beamtracking algorithms make directional antennas core aspects in 5G cellular networks and also certain Wi-Fi routers.
Incorporating realistic gain profiles into stochastic geometry models has been taxing towards mathematical tractability, and therefore sectorised approximations of the above are often used instead \cite{Bai2014a,Zhou2011} in order to derive closed form expressions for the network coverage, data rates, and multiuser transmission sum rates \cite{Bai2015}.
At such high frequencies however, blockage effects from pedestrians and cars become significant so several models have been proposed to capture these effects as well \cite{Venugopal2015,Bai2013}.

One of the latest advancements in network connectivity related to directional antennas is that of wireless power transfer.
These technologies pose many new opportunities and challenges towards the development of energy-neutral wireless communication networks \cite{Kansal2007}. 
One particularly interesting application is that of Simultaneous Wireless Information and Power Transfer (SWIPT) where the receiver may split its received signal into different domains, e.g., time, power, antennae, and space and process it accordingly \cite{Krikidis2014}.
Optimal power splitting strategies exist and depend on a number of factors including the deployment density of base stations, and their antenna directivity \cite{DiRenzo2017}.
Namely, SWIPT networks with directional antennas are generally more robust with respect to both information and power coverage \cite{Georgiou2017}.

\section{Temporal networks}
Having focused so far on the connectivity behaviour of SRGGs under various limits,  we now turn our attention to temporal networks.  In our context, this refers to networks
with a fixed set of nodes together with links that change with time.  As always, we are interested in spatial networks and the effects of spatial structure and boundaries.
We explore network properties such as the expected delay (the time it takes for a node to make a link), or the minimum time for paths to form .
Even when ignoring any underlying spatial geometry of the network it is often difficult to provide closed form expressions for things such as path formation. 
Recently, \cite{tajbakhsh2017accessibility} provided upper and lower bounds for the probability of accessibility (probability there is a path between i and j at time t)  of a network for the general case when links between nodes are random, and possibly time dependent, and is only tractable when the link probabilities are identical across the network.
Interestingly, their predictions for the accessibility probability perform well when compared with the inter-contact time of Taxis in Rome, where taxis are said to be connected if they are within some critical distance \cite{bracciale2014crawdad, tajbakhsh2017accessibility}. 
By modelling the probability a link is made in a given time slot by an exponential random variable they are able to capture the characteristics of a temporal-spatial network.  
For further discussions on space free temporal networks,the reader is referred to \cite{holme2012temporal} for a review while \cite{boccaletti2014structure} provides a thorough overview of dynamics on multi-layer networks.

Incorporating the spatial structure of wireless networks naturally increases the complexity of the analysis.
One solution is to model the network dynamics by fixing the underlying structure of the point process $\Phi$ and only allowing for the set of edges to vary with time.
Indeed, by this model \cite{DG17} obtained closed form expressions for the probability the network is fully connected as a function of time by analysing  the distribution of isolated nodes for a uniform PPP on the torus (enabling $\Bar{N} < \infty$, but ignoring boundary effects) where the pairwise link probability depends on their Euclidean separation.
In these static networks, it is again those nodes that are highly isolated that hinder the flow of information through the network but can be improved if a random re-wiring of the network is done.

An interesting variation of this model is when an ALOHA channel access scheme is employed, in this scenario a node can either transmit or receive (half-duplex) a message during each time step.
This model adds directionality to the network  where the possible edge set varies with time.
By considering a connection function where two nodes connect if they form a receiver-transmitter pair, a noise condition is met and  there is no intermediary node that is transmitting, \cite{ganti2009bounds} showed that the time for a path to form between a source and destination scales linearly with their Euclidean separation. 
Moreover, \cite{baccelli2010new} highlighted that for the SIR model there is a phase transition for a critical transmit probability $\wp$ where the mean delay becomes infinite.  
This work was extended to nearest neighbour communication models by \cite{haenggi2013local}, where they also provide bounds on the delay of Poisson networks. 
An infinite mean delay is a consequence of there being arbitrarily large voids in the Poisson network, so conditioning on there being another point in the process mitigates this. However, even conditioning on there being two points in $\Phi$, the expected shortest delay between the points grows faster than their Euclidean separation \cite{baccelli2011optimal} .   

It is often convenient to assume an infinite mobility model in the network where there is a new, independent, realisation of $\Phi$ at each time slot, i.e. there is no spatial correlation between time slots, and as a consequence simplifies the analysis. 
By employing this method, coupled with the static case, one can obtain upper and lower bounds for the performance of these spatial-temporal networks with mobility (see Sec.\ref{s:Mobility}).
For the high mobility case the local delay is always finite for the SIR model \cite{haenggi2013local} due to lack of correlation between time slots, this alludes to how mobile networks have the potential to resolve problems of disconnectivity.

\section{Mobility}\label{s:Mobility}
\subsection{Mobile networks}
In mobile networks there is no fixed network topology, instead, nodes move around the domain according to a particular set of rules.
This resultant mobility causes links to be continually made and broken. 
Wireless communication networks are a natural application where the nodes could represent hand held smart devices or vehicles say.  
Of particular interest are decentralised mobile (ad hoc) networks since as the number of smart devices continues to grow, so does the strain on the pre-existing network architecture. 
By relaying packets in a multi-hop fashion, rather than through a centralised router, the network becomes easily scalable without large overheads \cite{Helen2014}, provided the devices are mobile.
The importance of mobility to ad hoc network performance was highlighted by both \cite{Gupta2000} and \cite{Grossglauser2001}.
For the static case, comprised of $n$ nodes with fixed transmit power, Gupta and Kumar \cite{Gupta2000} showed the capacity per node of the network scales like $O\left(\sqrt{\frac{1}{n\log n}}\right)$, suggesting network performance decreases with node density.
However, Grossglauser and Tse \cite{Grossglauser2001} showed that in an interference limited environment mobility can in fact improve network capacity; albeit at the cost of increased delay.
As one might expect, network performance remains sensitive to the the choice of mobility model used, for instance in \cite{Lin2004, Sharma2004, Neely2005} showed that the delay-capacity trade-off differs for the random waypoint and Brownian motion models (see below), and thus characterising the level of inhomogeneity is important\cite{Schilcher2017}. 
For the remainder of this section we discuss a number of interesting and practical mobility models.

\subsection{Random walk}
One of the simplest mobility models is the random walk (RW) where nodes move independently from one another, and their direction of movement at each time step $T$ is chosen at random; thus a particular nodes location at any time $t$ is simply $\bx(t) = ( t - \lfloor\frac{t}{T}\rfloor) \bv_{\lfloor\frac{t}{T}\rfloor} + \sum_{i = 0}^{\lfloor \frac{t}{T}\rfloor - 1} \bv_i T$,  with $v_i$ denoting the velocity at time $i$.
In finite domains, the path is reflected off the boundary and the resulting spatial node distribution is uniform \cite{Bandyopadhyay2007}; for dim $\le 2$ the RW is recurrent.  
As a consequence key metrics are often analysed using a uniform point process \cite{Gong2014} and results are compared with other models that have an asymptotic stationary distribution, but ignores any inhomogeneities in the network.
Alternatively, it is sometimes convenient to consider a mobility model on a lattice, where the vertices represent intersections of streets in cities such as New York; one such model is the correlated RW, which is a generalised version of the standard RW\cite{Bandyopadhyay2007}. On the two dimensional lattice a user continues in the same direction  with probability $\wp$, opposite direction with probability $q$ and orthogonal direction with probability $2r$, such that $\wp + q + 2 r = 1$.
A further extension of the RW is the Manhattan model where $q=0$, i.e you never revisit the last lattice site, and the speed between consecutive time steps and other users on the same street are correlated \cite{Bandyopadhyay2007}.  

\subsection{Random waypoint}\label{s:rw}
The next, and arguably most well studied of the mobility models, is the Random waypoint (RWP) mobility model, which has an asymptotic (non-uniform) stationary distribution. 
In the RWP node movements are independent from one another and a single node chooses a waypoint uniformly at random, travels to it with a constant speed, pauses for sometime with probability $\wp_T$ then repeats the process. 
The time a node waits at each waypoint, i.e. its ``think time", can be either be constant or vary from waypoint to waypoint depending on the model.
As such, the RWP can be characterised by a sequence of waypoints and pauses, and unlike the RW a node continues on a path often for multiple time slots.
Due to the continual crossing of paths in the middle of the domain, the probability of finding a node in the bulk is higher compared with the boundaries; an effect which is argued to capture the mobility patterns of users in a city.
The stationary distribution of the RWP has a simple closed form in 1-d \cite{BRS03} and an integral form for any convex polygon which is easily computed numerically \cite{HLV06}. 
Interestingly the spatial distribution of nodes
in the RWP model is exactly that of the betweenness centrality of a uniform network in the disk (and other convex domains), in the limit as the number of users $\to \infty$ \cite{GGD15}. 
Intuitively, this is a consequence of nodes within the bulk having increased importance as they are more likely to lie on the shortest multi-hop path between any pair of nodes.
The mobility of the RWP leads to the outage probability being both spatially, and temporally, correlated \cite{Koufos2016a,Gong2014} in an interference limited environment; an affect which increases in a dense network with blockages \cite{Koufos2017}.

\subsection{L\'evy}

A L\'evy  mobility model (sometimes referred to as a scale free RW \cite{Benhamou2007}) is a modified RW where the path lengths are taken from a heavy tailed distribution, thus having infinite second moment, meaning that long ``flights" occur with a power law frequency rather than being exponentially rare \cite{Mantegna1994}. 
These heavy tailed distributions are interesting in the analysis of wireless networks since they are also a characteristic of human mobility \cite{Benhamou2007, Brockmann2006, lee2011delay}, which was observed from the traces of bank notes \cite{Brockmann2006}. As such, L´evy mobility has been used to model the spread of infectious diseases due to air travel, and the mobility of portable smart devices in wireless networks \cite{lee2011delay}.

There are typically two cases studied: the L\'evy flight and L\'evy walk, where the former has each flight taking a fixed time, and the latter having finite velocity culminating in a strong spatial-temporal correlation \cite{lee2011delay}.
These flights of large length $l$ follow a power law distribution 
$f_X(l) \sim l^{-\alpha -1}$, with $ 0< \alpha < 2$, which exhibit in a self-similar manner resulting in a typical trajectory having a fractal dimension of $\alpha$ \cite{Hughes1995,Chechkin2006}. 
The scale free nature of L\'evy mobility models leads to a super-diffusive behaviour, but when think times are also modelled using a power law, the model can either be super-diffusive or sub-diffusive \cite{Rhee2011}.

Analogously to how the sum of i.i.d random variables with finite mean converges to a Gaussian under the CLT, the sum of these i.i.d random variables with infinite second moment tends to a symmetric stable L\'evy distribution law with density  \cite{Kolmogorov1968,Chechkin2008,Dettmann}
\begin{align*}
f_{stable}^{\alpha,c}(x) 
&= \frac{1}{2\pi}\int_{-\infty}^\infty \exp\left(- i t x  -|c t|^\alpha\right) \dd t 
\end{align*}
where $c>0$ is a scale factor. 
For $\alpha = 1$ it reduces to the Cauchy distribution, whilst the Gaussian distribution is recovered when $\alpha \to 2$.
For the L\'evy flight model Ref~\cite{lee2011delay} show that the critical delay behaves like $\bar{N}^\alpha$, whereas for the L\'evy walk   the delay is $\bar{N}^{\frac{1}{2}}$ for $\alpha <1$ and for $\alpha \ge 1$ it behaves like the L\'evy flight.
This transitional behaviour at $\alpha =1$ for the L\'evy walk is a consequence of the mean flight length being infinite for $\alpha < 1$ \cite{lee2011delay,lu2014scaling}

The truncated L\'evy flight was later introduced to ensure a finite second moment \cite{Shlesinger1986}.
Each flight has length $l$ chosen from a levy stable distribution, and is re-sampled if the length is less than zero or greater than some cut off length $l_{\text{max}}$. Similar to the normal levy flight model the direction of travel and speed are chosen uniformly from $U(0,2\pi)$ and $U(v_{\text{min}},v_{\text{max}})$ respectively; as such the mobility can be described by a sequence of flights and pauses. 
At each destination, the pause time is sampled from a different levy stable distribution and is re-sampled if  it is less than zero or greater the specified maximum time $t_{\text{max}}$.

\subsection{SLAW}
Arguably, the Self-similar least action walk (SLAW) model \cite{Lee2012} provides a more accurate model for human mobility, which was shown when they compared simulation to real life traces, but in contrast to those previously mentioned lacks a rigorous mathematical formulation. (You can download the simulation in a link provided in the paper).
It aims to capture the 4 key features of human mobility: flights and pause-times follow a truncated power-law; inter-contact times also follow a similar power-law decay; human mobility exhibits heterogeneous features and waypoints are fractal in nature. 
Essentially this model captures how humans continuously revisit the same places (work, home, gym etc) in their daily lives, which defines a concept of a local area of mobility, but they occasionally travel long distances (visit family, days out), whilst the places they do visit tend to be popular. 

More generally, \cite{Santi2005} studied the critical transmission range needed for the RGG for a general mobility $\mathcal{M}$ and in particular for the  RWP model showed $r_c$ is $O\left(\sqrt{\frac{\log n}{n}}\right)$ for a non-zero pause time. 
Their analysis holds more generally for any bounded mobility model without blockages. 

There is a plethora of other models which claim to capture at least one feature that characterises human mobility with the analysis of these models being largely focused on comparing them to real life traces and simulating network behaviour.
Therefore, for the application of wireless networks it largely remains  a balancing act between mathematical tractability and model accuracy.
One approach is to focus on particular aspects of human mobility such as regions of high/low densities or the fractal distribution of waypoints \cite{chen2018capacity, D17} that capture the essence of the problem and allow for the analysis of some key network metrics.

\section{Fractals}\label{s:fractal}
So far, we have considered only fairly simple domains.  The RGG and SRGG have mostly been studied on square or flat torus domains, with occasional forays into rectangular, more general polygonal and some examples with curved boundaries and/or obstacles.  However, both from a mathematical and a practical point of view, it is important to consider more complex geometries.  A good starting point for natural fractals was Richardson's observation that the length of rivers and coastlines depends on the length scale used to measure them with an exponent related to a non-integer fractal dimension\cite{Richardson61,Mandelbrot67}.   Fractals are important in biology, for example trees and lungs~\cite{NLW13}, where they solve optimisation problems such as maximal surface area for a given volume.  Both natural features and optimisation leads to fractal structures in the built environment, such as in land use and transport networks~\cite{Shen02}. For wireless applications specifically, Ref.~\cite{Geea16} observes from empirical measurements that the coverage domain of a cellular base station is fractal, and the popular self-similar least action walk (SLAW) model for human mobility uses fractal distributions~\cite{LHKRC09}.

Spatial networks with fractal boundaries were studied in Ref.~\cite{Dettmann2015}.  The fractals were defined using a self-similar construction: Let $\{T_i\}$ be a finite set of contracting similarity transformations, that is, $|T_i{\bf x}|=r_i|{\bf x}|$ with $r_i<1$. Then, there is a unique non-empty closed set $F$ satisfying $F=\cup_iT_i(F)$ as shown by Hutchinson~\cite{Hutchinson79}.  The open set condition says that there exists an open set $V$ with $\cup_iT_I(V)\subseteq V$.  In this case, the Hausdorff and Minkowski dimensions are both equal to the similarity dimension, which is the unique positive solution $D$ of $\sum_ir_i^D=1$.   It would be interesting to investigate fractal boundaries where these dimensions are not equal, such as some classes of self-affine sets~\cite{Falconer13}.  There are also some other mathematically natural random fractal constructions including critical percolation~\cite{BH12} and aggregate tessellations~\cite{TZ01}.

In Ref.~\cite{Dettmann2015} the connection rule was line of sight (LOS).  At high density, there are many nodes located near the boundary with a positive probability of not linking to the rest of the graph.  This leads to a connection probability of the form $\exp(-a\lambda^{D/d})$ where $a$ is a constant (or generally a log-periodic function of the density $\lambda$), $d=2$ is the dimension of the underlying space, and $1<D<2$ is the similarity dimension of the boundary.   In particular, the connection probability decreases with density, and in practical networks
would need additional nodes situated near the opening of small enclosures near the boundary.

An alternative scenario is for the measure defining the PPP to be fractal.  These were considered in Ref.~\cite{D17} along with some non-fractal self-similar measures.  Here, the defining transformation is $\Lambda=\sum_i p_i\Lambda\circ T_i^{-1}$ with $p_i$ a probability vector (that is, $p_i\in(0,1)$ with $\sum_ip_i=1$).  If $p_i=r_i^D$, the measure is ``almost uniform'' (AU) in the sense that there is a fixed $\epsilon>0$ for which any ball centred on a point in the support $F$ contains at least $\epsilon$ times the measure in any other ball centred on a point in $F$ and of the same radius. Ref.~\cite{D17} investigates whether the number of isolated nodes is Poisson distributed and whether the network is likely to connect if there are no isolated nodes.  The first property can fail if the measure is not AU, and the second can fail if the support is finitely ramified, that is, can be disconnected by removing a finite number of points.  Whether the measure is fractal or smooth is less important.

\section{Metadistributions}\label{s:meta}
In any system with a source of randomness other than the node locations, it is helpful to consider metadistributions, that is, the distribution of some quantity conditioned on fixing the node locations.  
The first example of such an analysis was for Poisson bipolar networks, where the set of transmitters are modelled by a PPP, and each transmitter is paired with a receiver located at fixed distance $R$ away in a direction chosen uniformly  at random~\cite{GA10}.
As a result, bipolar networks are comprised of transmitter and receiver pairs, and there no longer exists a mesh network structure where a device has the potential to form a connection between multiple devices (directly or indirectly in a multi-hop fashion.)
The authors in Ref~\cite{GA10} considered the metadistribution of the link outage probability, whilst Ref.~\cite{Haenggi16} which considered the metadistribution of the signal to interference ratio (SIR).   Each link in the network has an outage probability depending on the locations of the other nodes, thus, fixing a large network we find a distribution of SIR.  Since the PPP is an ergodic process, this also gives the distribution of outage probabilities for a transmitter located at the origin, considering randomly located nodes.  Averaging over the node locations, or equivalently, the metadistribution, we obtain the mean outage probability.  The motivation for
understanding the metadistribution is that it gives much more information than the mean about the performance of typical individual links.  

The SRGG has a source of randomness other than node locations, namely the links. Given the node locations, each node has an isolation probability, that is, the probability
that it has degree zero. Ref.~\cite{DG17} considered the metadistribution of the isolation probabilities.
In this paper, and previously in Refs.~\cite{GA10,Haenggi16} it was noted that the metadistribution is not generally available in closed form, however it is easier to find
analytic expressions for the moments using the probability generating functional.  Mnatsakanov's method~\cite{Mnat08} was used to extract the metadistribution numerically
from these moments.  For short ranged connection functions the metadistribution peaked at zero and/or one, whilst for a small world generalisation involving longer links, it
became more concentrated towards a value strictly between zero and one.  This paper also considered a temporal SRGG, with links correlated in time as determined by an
Ising spin model, using the above calculation to investigate the distribution of times required to send information to all nodes on the network.

\section{Entropy}\label{s:entropy}
Randomness can helpfully be understood using the notions of information theory and entropy.  Given a discrete probability distribution with probability mass function
$p_i\in[0,1]$ with $\sum_i p_i=1$, the Shannon entropy is well known to be $H=-\sum_i p_i\log p_i$ (the logarithm is often base 2, but may be $e$ or 10 and we adopt the convention $0\log 0 = 0$).
The concept of entropy appears in the
network literature in several distinct ways.  
The earliest use is to describe the information content of a single graph by constructing a probability distribution on the
nodes~\cite{Rashevsky55}; the many approaches along these lines are reviewed in Ref.~\cite{DM11}; see also the more recent Ref.~\cite{HOSVSZD13}.  
Other work
considers entropy of processes on the network, such as flow of water~\cite{AJ03} or nerve impulses~\cite{CMMBA09}.  Alternatively,
entropy can refer to a graph ensemble, a probability distribution on graphs~\cite{JBWT08}, for example arising from a statistical mechanical approach~\cite{Bianconi09}.

Most work to date have focused on non-spatial graph ensembles.  More recently, spatial graph ensembles have been considered; the spatial character of the graph leads to
distinctive properties of the entropy as a function of the parameters.   The first known work to allude to the entropy
of SRGG ensembles appeared in the wireless communications networking community~\cite{TBH05}.  Here, randomness may arise from both node locations and links; for the
RGG the only randomness is in the node locations, whilst for the SRGG it is interesting to consider entropy conditional on fixing the locations and retaining the random
links~\cite{HMB14}, in a similar vein to the metadistributions discussed above.  Averaging this entropy over the node locations~\cite{Coon16,CDG17,CS17} then gives the
conditional entropy, related to the mutual information between the graph topology $G$ and the node locations $P$ through the relation $I(G,P) = H(G) - H(G|P)$.

\section{Conclusion}
We have reviewed models for spatial networks,, for which there is a growing
literature in probability, statistical physics, complex networks and wireless communications, among other fields.  The original model is the random geometric graph (RGG), studied for almost 60 years, and of ongoing interest.

As we have seen, the RGG has many generalisations, to different point processes, linking rules as a function of distance and orientation, confined geometries (including non-convex), temporal and mobility effects, complex (fractal) geometry in the point process and/or boundary, and interference (multi-point interactions).  There are some instances of universality, for example, that the connectivity probability can be expressed in terms of boundary components and moments of the connection function, and is thus to a large extent independent of the details of the geometry and connection function. However, most generalisations have led to completely new qualitative behaviour.

The scope of open problems in this field is truly vast. Even an apparently simple question as giving an effective approximation for the connection probability of the soft random geometric graph in one dimension has not been solved.  Extensions and combinations of the above topics will provide a source of mathematically and practically interesting problems for many years.  Whatever the reader's inclination, the following questions may be helpfully posed: How do the results for spatial networks differ from non-spatial (for example random graph) models?  How do the results depend on details of the geometry (for example point processes, confining boundaries)? What are the properties of typical networks, avoiding averaging over locations and links?

\onecolumn{
\section*{Acknowledgements}
The authors would like to thank Justin Coon for helpful discussions and the anonymous reviewer for their insightful comments.
This work was supported by the EPSRC [grant number
EP/N002458/1].
The authors would also like to thank the directors of the Toshiba
Telecommunications Research Laboratory for their support.
In addition, Pete Pratt is partially supported
by an EPSRC Doctoral Training Account.
This paper contains no underlying data.
\bibliographystyle{custom3}
\bibliography{merge_edits}
}

\end{document}